\documentclass[manuscript,screen,authordraft,nonacm,review=false,timestamp=false]{acmart}
\AtBeginDocument{%
  }

\setcopyright{acmlicensed}
\copyrightyear{2025}
\acmYear{2025}
\acmDOI{XXXXXXX.XXXXXXX}
\acmConference[FAccT '25]{2025 ACM Conference on Fairness, Accountability, and Transparency}{June 23--26,
  2025}{Athens, Greece}
\acmISBN{978-1-4503-XXXX-X/2018/06}

\begin{document}

\title{Governing AI Beyond the Pretraining Frontier}

\author{Nicholas A. Caputo}
\email{nick.caputo@oxfordmartin.ox.ac.uk}
\orcid{0009-0006-9482-8563}
\authornotemark[1]
\affiliation{%
  \institution{Oxford Martin AI Governance Initiative}
  \city{Oxford}
  \country{UK}
}

\renewcommand{\shortauthors}{Caputo}

\begin{abstract}
  This year, jurisdictions worldwide—including the United States, the European Union, the United Kingdom, and China—are set to enact or revise laws governing frontier AI. Their efforts largely rely on the assumption that increasing model scale through pretraining is the path to more advanced AI capabilities. Yet growing evidence suggests that this “pretraining paradigm” may be hitting a wall and major AI companies are turning to alternative approaches, like inference-time “reasoning,” to boost capabilities instead.
  
  \begin{flushleft}
      This paradigm shift presents fundamental challenges for the frontier AI governance frameworks that target pretraining scale as a key bottleneck useful for monitoring, control, and exclusion, threatening to undermine this new legal order as it emerges. This essay seeks to identify these challenges and point to new paths forward for regulation. First, we examine the existing frontier AI regulatory regime and analyze some key traits and vulnerabilities. Second, we introduce the concept of the “pretraining frontier,” the capabilities threshold made possible by scaling up pretraining alone, and demonstrate how it could make the regulatory field more diffuse and complex and lead to new forms of competition. Third, we lay out a regulatory approach that focuses on increasing transparency and leveraging new natural technical bottlenecks to effectively oversee changing frontier AI development while minimizing regulatory burdens and protecting fundamental rights. Our analysis provides concrete mechanisms for governing frontier AI systems across diverse technical paradigms, offering policymakers tools for addressing both current and future regulatory challenges in frontier AI.
  \end{flushleft}

\end{abstract}



\received{22 January 2025}

\maketitle

\section{Introduction}
Frontier AI regulation stands at a turning point. In the United States, the Trump Administration just rescinded Executive Order 14110, the country’s main framework for regulating frontier AI\cite{Shepardson2025}, and may soon put forward its own approach. The European Union AI Act’s provisions on General-Purpose AI (GPAI) are set to come into force this summer\cite{EC2024} and the government of the United Kingdom has promised to pass a bill dedicated to frontier AI\cite{Gross2024} this year. China may promulgate its own comprehensive AI law, which has been circulated in draft since 2023\cite{CSET2024}. At the same time, international processes in forums like the United Nations\cite{UNAI2024}, OECD\cite{OECD2024}, and AI Summit Series\cite{DSIT2024} raise the prospect of a global regulatory framework.

These efforts reflect a worldwide consensus that frontier AI, “highly capable general-purpose AI models that can perform a wide variety of tasks and match or exceed the capabilities present in today’s most advanced models”\cite{DSIT2023}, requires serious regulatory treatment. Though far from the only form of AI in need of regulation\cite{mayson_bias, 10.1145/3630106.3658541}, frontier AI presents novel challenges to global safety and security\cite{Anderljung2023MER, Egan2023, Pouget2024} because of its broad and rapidly-improving capabilities. Governments are seeking to meet these challenges through regulation this year.

Unfortunately, nearly every major regulation in force or in draft relies heavily on a key technical assumption that has been undermined by recent events: the belief that “scaling” AI models through pretraining runs with ever more compute and data is the primary driver of frontier AI capability gains. This assumption shapes core regulatory mechanisms, from triggers for legal coverage(like the EU AI Act's 10\textsuperscript{25} FLOPs threshold\cite{EUAIAct51}) to enforcement strategies (as with US export controls on advanced microchips\cite{ExportCtrls2025FactSheet}). The assumption that scale is the main driver of capabilities has been true in the “pretraining paradigm” of the last several years\cite{Piper2024}, and this paradigm has had benefits for regulatory design because it has created conditions of relative transparency, predictability, and centralization in the field of frontier AI. Scaling requires massive and increasing quantities of scarce resources, particularly compute and energy, and regulators can track which companies are scaling pretraining by tracking those resources\cite{sastry2024computingpowergovernanceartificial}. The "scaling laws" that have governed this process also roughly predict the capabilities generated by a given run, so regulators can approximately predict what kind of advance a new model will be based on its inputs\cite{doi:10.1073/pnas.2311878121, hoffmann2022trainingcomputeoptimallargelanguage, kaplan2020scalinglawsneurallanguage}. Governments can then prepare for risks that might be created by these advances and focus regulation and evaluations on the large companies making them, avoiding mistargeting and keeping regulatory burdens away from users and on big companies that can best bear them\cite{sastry2024computingpowergovernanceartificial}.

But there is increasing evidence that the pretraining paradigm is ending, with serious consequences for regulations predicated on its continuing. Recent reporting suggests that leading frontier AI companies have struggled to build the next generation of models through further scaling pretraining\cite{information2024, bloomberg2024, wsj2024} (though there is disagreement here\cite{Clark2024, SemiAnalysis2024}), and some leading researchers agree, arguing that scaling pretraining is hitting a wall imposed by the limited supply of good training data\cite{Sutskever2024, Robinson2024, Narayanan2024}. Even if the "data wall" has not been hit yet, the exponential demands of scaling pretraining mean that, absent breakthroughs, resource constraints will begin to bite in the coming years\cite{villalobos2024rundatalimitsllm}.

At the same time, frontier capabilities continue to improve\cite{OAIo1} and companies are making huge investments betting progress will go on\cite{Duffy2025, Gardizy2024, Bajwa2024}. Recent breakthroughs in “reasoning” systems point to alternatives to scaling pretraining with new uses and even better benchmark scores\cite{OAIo1, Jones2024, Chollet2024}. This decoupling of scaling pretraining and capabilities improvements could cause serious problems for frontier AI laws premised upon their relationship continuing. If the pretraining paradigm ends but rapid AI progress endures, the new regulations being worked out around the globe will become misaligned with the technology’s trajectory and lose their effectiveness just at the moment that risk from new development pathways and increases in capabilities intensifies.

This paper examines how frontier AI regulation could adapt if the pretraining paradigm ends, providing both concrete recommendations for new paths forward and broader lessons for governing advancing technologies. We make several specific contributions:
\begin{itemize}
    \item We analyze current regulations to identify key vulnerabilities from their reliance on pretraining scaling. \item We introduce the concept of the “pretraining frontier,” the capabilities ceiling on scaling pretraining alone imposed by current resource constraints, and explore the implications for industry structure and regulation if it becomes significantly binding on capabilities growth. \item We discuss how efforts to move beyond the pretraining frontier will shape the behavior of frontier AI companies, creating regulatory challenges and opportunities. \item We examine alternatives to current regulatory approaches, focusing on increasing transparency; monitoring of inputs like data, inference compute, and algorithmic innovation; and enhancing regulatory capacity.
\end{itemize}

The pretraining paradigm has allowed regulation to be relatively focused and light-touch and to focus on well-resourced companies rather than users. Those virtues should be replicated in the new capabilities paradigm, balancing safety, respect for rights, and innovation. But the end of the pretraining paradigm means regulators will face a more complex environment with more actors pursuing diverse approaches to capabilities advancement and new risks emerging, increasing the chance that regulations are either ineffective or overbearing. Our analysis seeks to offer practical steps through this transition, though more work is needed to flesh out which paths to take. A key lesson of the paradigm shift we may be currently undergoing is that regulating the changing future is difficult, but the risks are real and regulation is happening now. Getting it right is crucial for ensuring that transformative AI capabilities develop in alignment with human values and interests.

\section{The Pretraining Paradigm}

Over the past few years, frontier AI has been driven by a simple principle: more scale means more capabilities. Researchers have found that pretraining autoregressive models with more and more compute and data produces better performance across a variety of domains without the need for architectural innovations or domain-specific engineering\cite{devlin2019bertpretrainingdeepbidirectional, brown2020languagemodelsfewshotlearners, Piper2024}. In many ways, this trend is just an instantiation of the “bitter lesson,” the general finding that AI capabilities growth mostly derives from increases in computing power rather than human breakthroughs\cite{Sutton2019}. But the development of the transformer architecture\cite{vaswani_attention_2017} and the marshaling of massive amounts of data and compute by frontier AI companies like OpenAI, Anthropic, and Google DeepMind has led to a step change in AI capabilities.

A key feature of this pretraining paradigm is its predictability. Researchers have documented the existence of “scaling laws” which show that cross-entropy loss (a measure of model performance) decreases reliably as data and compute increase such that the performance of the model being pretrained can be roughly forecasted from its inputs\cite{doi:10.1073/pnas.2311878121}. Cross-entropy loss correlates reasonably well with practical capabilities across a variety of economically valuable tasks, as an increasing number of use cases and benchmarks show\cite{reuel2024betterbenchassessingaibenchmarks, Mirowski, li2024pertevalunveilingrealknowledge, liu2024convbenchmultiturnconversationevaluation}. And because the resulting models are capable of aiding or performing economically valuable tasks like computer programming\cite{verge2024}, investment into capabilities will continue. The relationships among scaling, benchmarks, and the real world are not exact, and models have so far diffused into the economy slowly and underperformed expectations in many practical applications. At the same time, they also occasionally exceed expectations by displaying unexpected "emergent" abilities that were not predicted from their training objectives\cite{wei2022emergentabilitieslargelanguage}. But overall, companies have been able to forecast the returns to scale and make massive investments in improving frontier AI with relative confidence about what they're getting.

Importantly, though the basic lesson of the scaling laws is that more is better, certain ratios of compute and data inputs are optimal for getting the best performance from pretraining\cite{hoffmann2022trainingcomputeoptimallargelanguage}. Increasing one input beyond the bounds of optimality to compensate for limits on the other can provide moderate gains, but diminishing returns start to bite and the constraints become binding. For years, compute has been the limiting resource and companies have spent heavily on gathering enough of the scarce and powerful cutting-edge microchips necessary for large training runs\cite{Bajwa2024}. These training runs will, if scaling continues, only grow more massive, as firms push Moore's law\cite{CSIS2022} and invest billions of dollars into building out the compute and energy resources that will be necessary to support such runs.

Yet over the past few months, returns from scaling up pretraining seem to have begun to diminish. Many leading AI companies, including OpenAI, Anthropic, and Google DeepMind, have reportedly been disappointed with the models produced from their latest big pretraining runs\cite{information2024, bloomberg2024, wsj2024}. Eminent researchers, including some who have long trumpeted the benefits of scaling\cite{Sutskever2024}, have predicted the end of the pretraining paradigm, citing the limited supply of training data as the cause. Frontier models are mostly trained on large corpuses of internet text comprising trillions of tokens, small pieces of words, and the number of such tokens is finite. One estimate puts the total stock in the world at around 300 trillion, meaning that data was expected to run out in a few scaling generations if no data breakthrough is made\cite{villalobos2024rundatalimitsllm}, but leading models seem to use around 15 trillion tokens\cite{MetaLlama, DeepSeekv3} and companies are apparently running into trouble already. If frontier AI companies are already finding that the era of pretraining is ending, then it is possible that new directions will already have to be taken for AI progress to continue. And regulations premised on the idea that scaling with remain the driver of capabilities will have to change.

\section{Scaling-Based Regulation Around the World}

Despite frontier AI's mounting capabilities and risks, regulatory frameworks aimed at governing these systems remain relatively scarce, and many proposed laws have not made it to promulgation\cite{SB2024}. However, what regulations do exist have relied on the pretraining paradigm and would be undermined by its end. Analyzing how these laws operate today can inform how new, more resilient frameworks can be developed for the next paradigm of capabilities progress.

Before diving into key features of the regulations themselves, it is worth laying out a brief overview of some of their common aims and features. First, regulations aimed at frontier models generally address the risk that advanced AI systems will be capable of enabling or causing tremendous harms, particularly chemical, biological, radiological, nuclear (CBRN), and cyber- attacks, mass persuasion and disinformation campaigns, and loss of control of AI\cite{Anderljung2023MER}. Advocates of regulation, including from industry, admit that while these harms have not occurred yet, there is robust evidence to suggest that they are likely to manifest and at least a robust monitoring and oversight regime is necessary now to identify them when they emerge\cite{koessler2024riskthresholdsfrontierai, FMF2024}. Second, frontier regulations are often layered atop more targeted sectoral regulations in areas like housing and consumer protection that seek to prevent harms from AI like bias, discrimination, and exploitation occurring today\cite{EUAIActIII, Diberardino}. This layered approach provides a reasonable guarantee against the specific harms that only frontier models can cause, allowing sectoral regulators to focus on the areas that they are most competent in. Third, frontier regulators across jurisdictions rely on similar tools and on cooperation with each other to face this globally-emerging technology\cite{Cap2024}. Because of the nascent stage of both frontier AI and its governance, developing better understanding of frontier AI through scientific and evaluative tools and transparency and reporting requirements placed on model developers has been the focus of many regulations, though some specific obligations like risk tiering and associated mitigations have also been put into place\cite{EUAIActIII, EO14110}. Companies also have committed to a surprising degree of self-regulation\cite{FMF2024}. Finally, frontier regulations are generally aimed at natural bottlenecks like compute and at the large and well-resourced companies pushing the frontier of AI to reduce the cost and burdens of regulation by keeping them narrow and focused\cite{EUAIActIII, EO14110}. These initial steps are promising, but many rely on the pretraining paradigm as a trigger for coverage or to focus their application. The erosion of that paradigm risks seriously undermining their effectiveness.

\subsection{The EU AI Act and its Code of Practice}

The EU AI Act is probably the most significant and comprehensive AI law in the world and contains provisions dedicated to frontier AI. The Act classifies frontier AIs as "General-Purpose AI (GPAI) Models with Systemic Risk" and imposes special obligations on them\cite{EUAIAct51}. Providers of such models must evaluate the systemic risks generated by their models and act to mitigate those risks, as well as report serious incidents caused by the models and the steps taken to correct them\cite{EUAIActIII}. GPT-4 and similar models already qualify\cite{EP2024}. To guide companies in complying with the Act, the EU AI Office initiated the development of a GPAI Code of Practice (CoP), currently being developed\cite{EUCoP}. Frontier models will also be covered under the standard provisions of the EU AI Act with respect to their various applications, such as if they are used in critical infrastructure or to determine access to education\cite{EUAIActIII}. This two-tier system of regulation aims at preventing and mitigating harms up and down the stack of development and deployment.

The Act and CoP are intended to be future-proof and updated in line with changing technology. Article 51 of the Act, which provides the legal triggers for coverage as a GPAI with Systemic Risk, provides a useful demonstration of how the law seeks manifest this intent and also the difficulties of doing so. Article 51 provides two alternative triggers for classification of a model as a GPAI with Systemic Risk: first, evaluations that demonstrate that it has "high impact capabilities," and second, a decision of the Commission that it has such capabilities\cite{EUAIAct51}. Article 51 also includes a compute threshold provision set at 10\textsuperscript{25} training FLOPs, which creates the presumption that a model trained with more than that amount of compute is a General-Purpose AI Model with Systemic Risk\cite{EUAIAct51}. The draft CoP currently lays out some details for what is required for an evaluation to be good enough to be used to determine whether a model presents a given systemic risk and also lays out a taxonomy of what the "systemic risks" are\cite{EUCoP}. The Commission is required under Article 51 to update the coverage requirements to stay in line with advancing technology\cite{EUAIAct51}. 

Two problems have emerged from the decline of the pretraining paradigm that undermine these triggers. First, the law's language has already started to look outdated and a poor fit with new directions of the technology. The Act specifies that it covers "general-purpose AI model[s]," roughly those that "display[] significant generality and [are] capable of competently performing a wide range of distinct tasks"\cite{EUAIActIII}. It is already unclear exactly what is covered here. Is GPT-3 covered, for example? GPT-2? And the Act specifies that it covers "model[s]" and not "systems" or "AI" in general. This word choice, reasonable when the Act was drafted, is now proving a liability as "systems" and "agents" may replace "models" as significant sources of risk\cite{informationAgents, DeepSeekr1}, though the CoP is seeking to patch this gap\cite{EUCoP}. Second, the coverage triggers in Article 51, though written broadly, in practice will likely implicitly depend on having a small group of potential leading models that are known to the authorities that can be the focus of regulatory attention. As discussed, a model is covered if, one, it falls under the Act's definition of GPAI model quoted above, and two, it then meets either the evaluation or Commission decision trigger for GPAI model with Systemic Risk\cite{EUAIAct51}. But what counts as a GPAI model? And are all GPAI models subject to evaluations to see whether they present Systemic Risk? How will the EU AI Office allocate its scarce resources to cover this regulatory frontage? Article 51's 10\textsuperscript{25} FLOPs compute threshold thus will likely end up playing a more significant role than its language of "presumption" suggests because it provides an actual cutoff to answer these questions. But with the rise of reasoning models, that cutoff may not capture the real source of capabilities unless the language "cumulative compute used for [models'] training" in Article 51\cite{EUAIAct51} is stretched quite far. Given the difficulty of determining what the Act's coverage will be from its language alone, regulators have likely implicitly relied on there being a few leading companies on the frontier that they can track and focus on--any time OpenAI creates a new model, the Commission can decide it is covered. But if the pretraining paradigm breaks down, then, as discussed below, there will likely be many more companies producing frontier models and new sources of capabilities progress. In such a world, relying on a general sense of which models matter will not suffice to achieve the goals of the Act.

The Act and CoP are a promising step forward for regulating frontier AI and represent real progress in concretizing what frontier regulation will look like. Requirements like the Safety and Security Frameworks\cite{EUCoP} that mandate that companies investigate the risks posed by their models and lay out how they will mitigate them are useful steps forward and demonstrate practices that can be built on and adopted elsewhere. Furthermore, the various updating mechanisms embedded in the EU framework do go some way towards making it "future-proof"\cite{EUCoP, EUAIAct51} and adaptable to new generations of models. Regulating the future is difficult, and specificity in regulation is necessary to ensure rules mean something. But the effectiveness of regulation relies not just on the law and the regulators but also the deeper shifts of technology, economy, and society, and overspecification can prove a trap as those change.

\subsection{US Executive Orders and Export Controls}

The United States' frontier AI regulatory regime is being rewritten by the Trump Administration\cite{Shepardson2025}. But certain parts of the frontier AI framework which are vulnerable to the decline of the pretraining paradigm may persist in letter or in spirit into new regulations. The Biden-era frontier AI regulatory framework had two main elements, the Executive Order on the Safe, Secure, and Trustworthy Development and Use of Artificial Intelligence (EO 14110)\cite{EO14110} and various export controls placed on advanced microchips\cite{BarathChipsFall, BISDif}. EO 14110 was revoked by President Trump on his first day in office\cite{Shepardson2025}, but many of the rules that it called for are still in place pending rescission processes by relevant agencies and a new EO is likely forthcoming.

EO 14110, though itself revoked, may be the basis for future regulation and illustrates how the US government thinks of frontier AI regulation. EO 14110 defined frontier AI as being most importantly "dual-use" and included two key compute triggers for coverage that resemble those in the EU AI Act: One threshold for models trained with more than 10\textsuperscript{26} FLOPs (rather than the EU AI Act's 10\textsuperscript{25} FLOPs), and another for data centers with a theoretical maximum computing capacity greater than 10\textsuperscript{20} FLOPs\cite{EO14110}. Though these triggers were intended to be supplanted by ones defined by the Secretary of Commerce in consultation with others, they demonstrate the same reasoning and vulnerabilities to the end of the pretraining paradigm as the compute thresholds in the EU AI Act discussed above. The day after EO 14110 was released, the US established the US AI Safety Institute (AISI), inside of the National Institute for Standards and Technology, which seeks to improve the science of frontier AI and evaluate risks from models\cite{USAISI}. US AISI has so far mostly partnered with frontier companies to do evaluations, but its work forms the basis of future regulation.

The toothier part of American frontier AI regulation is the various export controls that have been put in place on advanced microchips. These export controls seek to ensure that the semiconductors necessary for massive AI training runs do not get into the hands of US adversaries, particularly China. Beginning in 2022, the Biden Administration began putting in place restrictions on what kinds of chips and design tools could be sold to Chinese entities\cite{BarathChipsFall}. Early this year, the Biden Administration released a major new set of controls dividing the world into three tiers, each of which is given different access to semiconductors, and banning the export of closed model weights to US adversaries\cite{HawkinsBloomberg}. In brief, countries in Tier 1, close US allies, face no restrictions on importing semiconductors, but countries in Tier 2, containing most of the world, are limited to importing 50,000 advanced GPUs per year unless entities in those countries conform to relatively onerous security requirements, in which case they can import more. Finally, countries in Tier 3 effectively cannot import advanced GPUs\cite{BISDif, CFRDif}. These diffusion controls seem to have survived into the Trump era.

These rules are premised upon the idea that scaling up AI training remains the key input for increasing model capabilities. If the pretraining paradigm breaks down such that inputs like algorithmic breakthroughs or larger supplies of data turn out to be the route to future progress instead, then restrictions on the export and use of GPUs will be useless because GPUs will not be the essential input for capabilities. Large training runs will likely still be necessary for progress in AI, but the extent to which they are the key chokepoint has become a question. And if, as discussed below, semiconductors become relatively plentiful because demand for them declines, then preventing Tier 2 and 3 countries from accessing them may become impossible, rendering the rules useless and only alienating other countries.

\subsection{The Forthcoming UK Frontier AI Bill and Existing AI Safety Institute}

The UK government has indicated that it will put forward a narrow bill regulating frontier AI this year, building on existing infrastructure like the UK AISI\cite{Gross2024}. While it is unclear exactly what the bill will contain, it will need some kind of trigger for legal coverage of AI systems that allows it to be targeted at frontier AI. The chosen trigger may resemble the compute thresholds discussed above with respect to the EU AI Act and US EO 14110 or it may point toward a new way to cover models as the pretraining paradigm declines and compute thresholds become less effective tools. Much more will be known (and can be analyzed) when the bill is released in the coming weeks.

However, the UK has made significant contributions to the international frontier AI governance regime already, particularly in the form of the UK AISI. UK AISI operates as a government-funded scientific authority that provides evaluations of frontier AI capabilities and pushes the science necessary for regulation forward\cite{DonelanAISI}. The AISI aims to fill the gap left by private frontier companies and evaluators, for example by open sourcing an evaluation platform that can be used to check for frontier model risks\cite{UK_AI_Safety_Institute_Inspect_AI_Framework_2024}. If the pretraining paradigm fails, the AISI's priorities and approach will continue to be useful, but there is some risk that it will be overwhelmed by the increased demands of dealing with a more diffuse and chaotic frontier. Relying on the continued concentration of frontier progress in a handful of companies could result in a situation in which AISI is unable to provide sufficient evaluation expertise to cover the frontier, and it would have to significantly grow to respond to this new challenge. Currently, the UK's relative lack of specific binding regulations and emphasis on capacity- and relationship- building has let it avoid many hard problems of technological change. But if it does take the promised steps to create more significant regulations this year, those regulations will have to be shaped to respond to frontier AI as it advances.

\subsection{Chinese Rules from Specific to Comprehensive}

China has a suite of laws aimed at regulating AI systems, but few of them so far have been aimed at frontier AI itself\cite{SheehanSafety, SheehanRoots}. The existing Chinese laws and regulations most relevant to frontier AI generally focus on specific applications of models, such as provisions governing algorithmic recommendation systems on the internet\cite{CLTAlgos} and provisions governing “deep synthesis” (effectively deepfake) algorithms\cite{CLTSynth}. The Chinese government has also sought to use these laws to build regulatory capacity and institutions over time, for example creating an "Algorithm Registry" in 2021 for an expanding list of kinds of algorithms now including frontier AIs\cite{CLTAlgos, SheehanSafety}.

As cutting-edge AI has shifted away from specific applications and toward the more general capabilities characteristic of frontier models, Chinese governance seems to be adapting toward a more general and full-stack form of regulation. The 2023 Measures for Generative AI, prompted by ChatGPT, aim not just at the use of models but also at their training and the data used to create them\cite{CLTGen}. In 2023, the Chinese Academy of Social Sciences circulated a draft Artificial Intelligence Law that would provide a relatively comprehensive overall regulatory regime and also included provisions targeted at frontier AI\cite{DigiCNAILaw}. The draft law points toward a higher-level, coordinated form of AI regulation with stronger teeth. Of particular note is the proposed Negative List, which sets out a list of kinds of models that must receive government permission before they can be deployed. There are also provisions aimed directly at frontier models that create enhanced reporting requirements for the development and use of these models. The draft National AI Laws requires that AI companies promote the safety (or security) of their AI systems, though exactly what that will entail remains to be determined at this point\cite{SheehanSafety}. Though the official version of the law has not been released as of it, it is likely that something will happen with overall regulation soon.

The Chinese regulatory regime may be best equipped to deal with a transition away from the pretraining paradigm. China has developed significant regulatory capacity in AI across a range of fronts, from narrow recommendation algorithms to leading foundation models like those developed by DeepSeek and Alibaba\cite{DeepSeekv3, qwen2}. In particular, the Algorithm Registry suggests that the Chinese government thinks that it can handle tracking and evaluating huge numbers of algorithms across a range of tasks, something that might be necessary if more companies can produce frontier models. China could also benefit from a shift away from compute as the binding constraint on capabilities progress because it would negate the effects of US export controls and shift the competition to different ground, where companies like DeepSeek could compete to push the frontier less handicapped by export controls. A more diffuse and decentralized frontier may not present the same challenge to Chinese authorities more used to monitoring a broad range of entities than those in the US or Europe. But exactly what path China will take will be unclear until a formal law governing frontier AI is put forward and its provisions clarified.

\section{What Comes After the Pretraining Paradigm?}

The pretraining paradigm may be coming to an end, but AI capabilities continue to advance rapidly. Companies are plowing hundreds of billions of dollars into AI capital expenditures, suggesting strong confidence in continued progress\cite{Duffy2025, Gardizy2024, Bajwa2024}. New research directions like the recently-released “reasoning” systems \cite{NatoO3, OAIo1, DeepSeekr1} demonstrate paths beyond pure scaling of pretraining. Especially if systems begin to build on each other, as may be happening with reasoning models using synthetic data generated by pretrained models\cite{NatoO3}, progress may be rapid.

Many futures for frontier AI are possible from here. Progress could stall into another “AI winter” characterized by slow growth, broken promises of global transformation, and backlash against recent hype. Technical breakthroughs could easily overcome the "data wall" and other obstacles to scaling up pretraining such that it remains the dominant source of AI progress. In either of these scenarios, the effectiveness of existing regulatory frameworks would likely be preserved because, in the first case, there would be few new frontiers to govern and, in the second case, existing approaches aimed at compute governance would continue to work. However, recent breakthroughs make change seem likely, creating a need for analysis of how frontier AI regulation could function in a new capabilities paradigm.

Three key variables could shape this potential new era: \begin{itemize}
    \item First, how far beyond current capabilities can companies push using new approaches? If further progress is easy, then incumbents will likely maintain their leads, but other companies might catch up if it is hard.\item Second, what is the main driver of capabilities? If human-driven innovation is the source of growth and, consequently, is the constraint on it, regulation may need to shift from focusing on predictable and trackable inputs like compute to more complex oversight of research and development.\item Third, how long will the new paradigm last? Will it be a brief interlude before some form of scaling returns? Or will it be a longer period of distributed growth?
\end{itemize}  

\subsection{A Crowded Pretraining Frontier?}

One crucial structural question will be whether the end of the pretraining paradigm creates a kind of cap on general capabilities progress that companies must use other approaches to get beyond. In such a world, a "pretraining frontier" might emerge at the maximum overall capabilities threshold enabled by scaling up pretraining. Recent reporting that OpenAI, Anthropic, and Google DeepMind have been unable to keep scaling model pretraining \cite{information2024, bloomberg2024} and the clustering of top models around the GPT-4 level over the past few years suggest that such a cap might exist. If alternatives like reasoning are not able to provide substantial increases in overall capabilities but rather push forward only parts of the capabilities graph (as some benchmarking of the reasoning models suggests\cite{NarayananBench, Narayanan2024}), then overall progress may slow and the pretraining paradigm's advantages for governance would dissipate. Leading companies may maintain scale and resource advantages to dominate the coming paradigm, a possibility analyzed below, but a shift toward deconcentration would have serious implications for governance and is worth discussing.

Slowed frontier progress could have significant structural effects on the industry. A lack of overall progress from the incumbents would likely mean that more companies could catch up to leaders, reaching the pretraining frontier and competing for breakthroughs beyond it. As the history of the last few years shows, there is only a lag of a few months or years between when the leading companies release a model and when other groups, including open-source developers, catch up and release a model with equivalent capabilities\cite{DeepSeekv3, DeepSeekr1, grattafiori2024llama3herdmodels}.

Furthermore, we should expect the pretraining frontier to get populated quickly if further scaling up pretraining slows. Lack of access to large quantities of compute has been a main obstacle to new companies joining the frontier AI race, and regulations like the US export controls have sought to leverage that chip scarcity to block competitors\cite{BarathChipsFall}. But access to microchips will become easier if the pretraining frontier is a hard cap. In such a world, big companies will no longer need to consume as much of the supply of the scarce microchips that are best suited for pretraining because they will already have reached the frontier. Combining this reduction in demand with Moore’s law, which increases effective compute supply\cite{CSIS2022}, the cost of enough compute to get models to the pretraining frontier may fall quickly. New companies could take advantage of this reduced cost to build models at enough scale to get to the pretraining frontier. Researchers at incumbent companies might leave to start their own labs because they see a chance to push ahead. An industry that has been relatively concentrated into a few key players around the world could deconcentrate as the resource moat that has prevented new entrants weakens and competition comes in.

The funding model for the AI industry might also change. In the current paradigm, the leading frontier companies are funded mostly by other giants\cite{Duffy2025, Bajwa2024, Gardizy2024}. A big company can justify investing billions of dollars in a frontier AI developer because scaling laws have made returns on that investment relatively predictable. But if the industry transitions to a mode in which there are many frontier companies competing to produce uncertain breakthroughs, it might be more difficult for big companies to justify putting their eggs in one frontier basket. Something more like the normal venture capital market dynamic, in which funders invest in a broad portfolio of moonshot frontier firms, might manifest, shifting the balance of power away from the incumbents and toward a broader set of frontier developers and funders.

Deconcentration may not occur because a new scarce good, for example chips specialized for inference compute, quickly replaces pretraining chips as the new bottleneck for industry players and regulators to target and again provides a significant advantage to the big incumbent players. Inference scaling suggests one such path forward\cite{NatoO3}, though OpenAI is already facing competition there from DeepSeek and Google DeepMind\cite{DeepSeekr1, RothGemini}. But until such a bottleneck clearly presents itself, and for as long as capabilities progress can come from a variety of sources, more and more companies may find the frontier of capabilities enabled by scaling pretraining. A deconcentrated and complex industry would be more difficult for regulators to monitor and oversee, increasing costs and the chance that risks emerge without regulatory coverage. Existing approaches that rely on concentration of industry and scarce compute would be less effective if those two conditions dissipate, as they might if the pretraining paradigm fails.

\subsection{A Return to Scale?}

But even a hard capabilities cap, if one emerges, will not last forever, and frontier AI companies are already finding ways beyond pretraining to improve capabilities. Many of these efforts will likely involve seeking new forms of scaling because scaling creates useful predictability for incumbent players and allows them to leverage their leads and resource advantages. The development of the new reasoning systems by OpenAI illustrates this trend. First, the reasoning systems seem to use inference scaling, spending more compute at inference time rather than during pretraining, to get better results\cite{NatoO3}. The continued success of inference scaling would mean that players with significant compute resources would retain an advantage. Second, leading companies are rumored to be using synthetic data generated by foundation models to train their new systems; synthetic reasoning traces generated in this manner are apparently a key part of reasoning systems\cite{NatoO3, DeepSeekr1}. This method transforms data into a new kind of compute scaling where companies can leverage existing compute resources to generate more and better synthetic data, improving capabilities and thus maintaining the advantage provided by their hyperscaler infrastructures. 

More broadly, companies may be seeking a kind of innovation scaling. If innovation is the key to further frontier AI progress, regularizing it such that it happens relatively predictably based on compute inputs would transform uncertain breakthroughs into forecastable progress. Automated R\&D is one way to push in the direction of regularized innovation. The new reasoning systems seem upon initial benchmarking to be particularly well-tuned for coding and less of a step up from prior models in fuzzier tasks like writing\cite{NarayananBench}. Some of that gap is likely a result of coding providing better ground truth to train systems on\cite{Narayanan2024}, but it is also probable that companies like OpenAI are seeking to automate increasing amounts of AI R\&D and improving the coding performance of their systems is the best way to accomplish that. Sufficiently automated research could look something like scaling, where companies can spend compute to increase innovation in a relatively predictable way, transforming breakthroughs into a Moore's law-style dynamic where progress shows a clean line on the graph despite depending on humans making apparently unpredictable breakthroughs over time. It seems likely that some company will break through the pretraining frontier in a way that makes rapid, regular capabilities growth possible again, but difficult to predict when this will happen. Figuring out how to regulate the interim and that new shift, when it comes, is the key task of frontier regulation.

\section{Regulating Beyond the Pretraining Frontier}

If the pretraining paradigm does give way to some new source of AI capabilities, how can frontier AI governance adapt? The pretraining paradigm provided two core advantages for governance, legibility and efficiency, that will likely diminish if the paradigm ends. Because of the predictability of scaling laws and the key position held by compute in capabilities progress, regulators had good visibility into the sources of risk and could target narrow but effective regulations at the entities best situated to bear the costs of regulation. These entities were usually big companies that could be regulated high up the AI stack, avoiding burdening the rights of users and intruding on their uses of AI. If the frontier field becomes more complex and risks increase, regulators will have to figure out how to address those risks while maintaining respect for rights and the light touch that has facilitated innovation so far.

There are two main ways that the regulatory field could become more complex if the pretraining paradigm declines: first, the number of companies competing to push forward the frontier substantially increases and second, the new capabilities approaches that they seek are less predictable and more risky than past methods. To confront these enhanced difficulties while preserving the virtues of legible and efficient regulation, regulators have three basic options. First, they could seek to increase the transparency of the frontier AI field and improve their understanding of it. Second, they could try to find or create bottlenecks that can be targeted to allow for more effective and less intrusive regulation, replicating the role of compute in the pretraining-oriented regulatory regime. Third, they could build up greater regulatory capacity to handle a more unpredictable and complicated frontier AI environment.

One of the main lessons of the paradigm shift that is likely occurring right now is that regulating prematurely and heavy-handedly based on technological assumptions that can later change is an ineffective recipe for governance. And each of the approaches just presented has its potential overall downsides. Trying to retain or recreate bottlenecks on the development of frontier AI will limit the development of these systems, locking away benefits that they could provide, and could artificially increase the concentration of the industry and provide significant opportunities for rent seeking and regulatory capture. Transparency and regulatory capacity building are a good and necessary steps for understanding and responding to the challenges of new technologies, but capacity can be used for bad ends and if the government is empowered to granularly monitor and control the use of a technology that will likely soon become an essential part of how people act in the world and develop themselves, abuses of rights could proliferate.

As risks from frontier AI increase and the deficiencies of current regulations become clear, calls for regulation will mount. This section analyzes how regulators could begin to respond to the breakdown of the pretraining paradigm in ways that avoid the worst traps of stifling or abusive regulation while ensuring the safety of their people.

\subsection{Transparency and the Science of Frontier AI}

The first requirement of effective regulation is getting an accurate understanding of the thing that is being regulated. Transparency into the operations of frontier AI companies is one key way to get that kind of understanding\cite{BallKoko, Ball}. Existing regulations have sought to improve regulators' view into leading labs\cite{EUAIActIII, EO14110} and benchmarking and evaluations help ascertain the capabilities of AI models such that companies and governments can determine what risks they pose and how to respond to those risks. These tools and frameworks provide a useful baseline for frontier AI regulation and increasing investment in them and speeding their rollout is useful.

However, the end of the pretraining paradigm creates challenges for transparency because of the potential increased complexity of the industry and of the sources of capabilities and risk. Instead of being able to concentrate regulatory resources on a small number of leading companies and provide robust evaluations and analyses of the top models, regulators may have to investigate more widely. 

The regulatory challenge will depend on how the end of the pretraining paradigm affects the frontier AI field. If a few large companies remain the leading players and can roughly predict the trajectory of capabilities improvements, then regulators will be able to continue targeting their attention on those companies and the new breakthroughs and approaches they develop. On the other hand, if the industry deconcentrates and many smaller companies in different jurisdictions can get to the capabilities frontier and push it in different ways, then transparency will be harder because regulators will struggle to cover the regulatory frontage. In such a world, regulators should try to improve their regulatory coverage as much as possible by seeking transparency into likely leading players and, in areas in which full coverage of companies is impossible, trying to cover sources of progress rather than players themselves. The number of paths forward for capabilities will likely be smaller than the number of companies at the frontier, as demonstrated by the return of reinforcement learning in the reasoning models\cite{NatoO3}. If regulators can get a sense of what the state of the art is in a given area of frontier AI, they can then monitor for overall breakthroughs in that area rather than looking at every specific company who might be contributing to progress. Similarly, grouping models by some criterion like parameters, overall risks, or evaluations scores would allow for economization on regulatory coverage because models that are similar across relevant criteria could be monitored as a group rather than individually. 

There are weaknesses to this kind of classification approach to transparency. Fundamentally, it relies on the relative predictability of sources of risk and the ability to identify them ahead of development or deployment. But if future architectural breakthroughs improve the capabilities of smaller models beyond a point at which they are capable of doing serious harm or a similar paradigm in which risk sources are very hard to predict, transparency into a select number of companies or models will be insufficient and increasing monitoring the only option.

\subsection{Regulating Data}

Regulators could also seek natural bottlenecks on capabilities progress that allow for focused attention like compute bottlenecks have allowed in the pretraining paradigm. Given that pretraining paradigm may come to an end because limitations on the supply of data prevent further effective scaling up of the size of models, data could become the new constraint on model progress that allows for targeted regulation. The use of synthetic data in recent reasoning breakthroughs\cite{NatoO3} makes this unlikely to be a panacea, but it is possible that access to key data sources will be necessary to expand capabilities within specific domains. If so, regulation targeting those data sources could have significant leverage.

Such regulation could broadly happen in two forms: First, as a kind of trigger for legal coverage (much like current compute thresholds\cite{EUAIAct51, EO14110}), and second as a point that can be regulated itself. In the first form, companies seeking to train models using more than some specified amount of data or perhaps using data of particular sensitive kinds would have to report that they were doing so to the government and the AIs that they created would be subject to capabilities or other kinds of evaluations. For example, models specialized through data selection for biological research might face more stringent requirements than other models because of the risks from biological uplift capabilities. In the second form, certain kinds of sensitive data would be held by the government or a private group such that they could be accessed only by vetted researchers. Ideally, some form of tracking of the data's use and distribution would also be developed such that the data would not leak without some ability to determine who the leaker is. Existing infrastructure around sensitive nuclear and biological research suggests that some version of this might be possible, and some civil society organizations have begun developing visions of responsible data stewardship that might involve creating restrictions on how data they manage can be used to ensure it is pro-social\cite{LeppertIDI}.

Data is unlikely to be the main obstacle to continued frontier AI progress, especially as synthetic data approaches begin to prove out. But if specialized data sources end up being essential to the development of certain risky capabilities, then developing monitoring or controlled access regulations aimed at those data sources might be a key step to understanding and mitigating risks.

\subsection{Regulating Overall Compute or Inference}

Compute may remain a useful target for regulation after the pretraining paradigm if it ends up being necessary for the success of new approaches like inference scaling. However, existing laws would have to expand their reach to cover forms of compute usage beyond pretraining. Furthermore, such expanded compute governance approaches would likely face significant new technical challenges and advances in distributed computing\cite{peng2024demodecoupledmomentumoptimization, intellect2024} suggest that they are unlikely to work as effectively as pretraining compute governance can today.

Regulating compute broadly, rather than in specific uses, seems like one place to start. In this kind of approach, rather than counting only the amount of compute used in pretraining a model, as the EU AI Act seems to today\cite{EUAIAct51}, all compute used in a model's development and deployment would be counted toward determining whether it qualifies as a frontier model and is subject to the relevant requirements for those models\cite{BrundageX}. Because inference in closed models happens in the cloud, cloud service providers like Amazon Web Services or Microsoft Azure could be made responsible for tracking when significant expenditures of inference compute were being made that hit a certain threshold above the cumulative compute that the model being used had been trained on. Unless the compute expenditure was authorized or being made by a trusted party, these providers would then report those expenditures to regulators who could investigate the usage to determine if it constituted a source of risk. Such a system would be more intrusive than the kind of simple compute governance that focuses only on whether a large training run is occurring because it would require learning something about what the compute was being used for. It also might not work technically because it could turn out to be possible to achieve dangerous capabilities in models without spending extraordinary amounts of compute such that such activity was differentiable for safe use. But if risks escalate sharply in the new paradigm, looking into more invasive forms of monitoring that still limit surveillance harms might be a necessary step to take.

It seems likely that some form of broader compute governance will be necessary in the forthcoming regime because spending large quantities of compute is a reasonably reliable source of capabilities improvements across domains. Furthermore, depending on the course of the technology, it is possible that a new form of compute governance that is less invasive than monitoring all large clusters and outputs will become possible. Developing the capacity to do such monitoring is a key step in technical governance that is only becoming more important as the technology of frontier AI shifts.

\subsection{Regulating Information}

Another potential target of regulation could be information in the form of algorithmic breakthroughs and similar drivers of capabilities progress. If it turns out that innovation is the key input for new forms of frontier AI, then the dissemination of the ideas behind these innovations will be a key bottleneck on how quickly different companies can catch up to and push the frontier. Furthermore, regulating algorithmic breakthroughs would allow governments to avoid regulating users and uses of frontier AI, instead focusing higher up the stack. Restrictions on classified information and the nuclear "born secret" regimes\cite{Morland_born} demonstrate that regulating information is possible in certain circumstances, though unlikely to be robust over time.

In general, regulating information in the form of algorithms or other kinds of breakthroughs would be unlikely to work for very long and probably poses costs too high to be warranted. Regulating ideas is extremely difficult, would likely stifle important benefits of AI progress, and also presents extreme risks of abuse. In a more diffuse and complex capabilities paradigm, it would be difficult for regulators to even know what information to regulate, and requiring researchers to register potential breakthrough ideas with the government for monitoring and processing would create a significant burden on innovation and provide uncertain benefits. Furthermore, information on AI improvements and how they happen seems to leak out quickly and other researchers have been able to replicate breakthroughs given little information about how they work. The replication of OpenAI's o1 with DeepSeek's R-1 within a few months\cite{DeepSeekr1, NatoR1} demonstrates the difficulty of preventing outside groups from figuring out how an advance works and finding a way to replicate it. Regulating information through classification and restriction might be necessary in some extremely narrow and dangerous domains, but it should not be a main strategy for regulating frontier AI.

However, it is possible that intellectual property (IP) law, a different kind of information regulation from classification and other control regimes, might become more important in a world in which innovations are the source of progress. Leading companies are already releasing fewer public research papers on their improvements\cite{BIPub} and they could seek to hold their innovations tighter still if they matter more for capabilities growth. IP law tries to strike a careful balance in encouraging innovation by rewarding people for their work while avoiding creating detrimental restraints on competition, and it is possible that it contains tools like patent buyouts\cite{Kremer} that would allow regulators to get more control over particularly dangerous advances while still rewarding innovation.

\subsection{Capacity-Building}

Finally, building up regulatory capacity would broadly benefit efforts to respond to changing technology by increasing the ability of regulators to understand where the technology is going and what the sources of risk are, both from new capabilities and from new actors. Existing investments into capacity-building like that into the UK AISI\cite{DonelanAISI} provide a useful foundation for progress here, but the United States in particular must ensure that it develops the capacity to keep up with the changing technology. Beyond generally gathering technical expertise, regulators should focus on forecasting how the technology could change and developing evaluation and monitoring systems that are flexible enough to manage different modes of progress. For example, many existing evaluations of frontier models focus on eliciting model capabilities (often through benchmark scores) rather than directly determining the relationship between those capabilities and the risks they present. Much cutting-edge evaluations work will always happen within leading companies as they try to understand the capabilities of the AIs that they build, but building out more sophisticated set of evaluations practices in government to test each part of the risk chain for frontier AI is necessary for regulators to have the ability to audit and effectively oversee the companies.

More regulatory capacity would also enable regulators to respond to a more complex frontier AI environment if the industry deconcentrates or if capabilities growth becomes less predictable. In either of these cases, regulators would have to cover a broader set of actors and risk sources than currently, which could present challenges unless they are able to get more resources. Focusing regulatory coverage by classifying groups of models and risk sources, as discussed above in relation to transparency, could provide one way to mitigate the overall increase in the cost of regulation, as could developing more sophisticated forms of automated evaluations that could be easily run on new systems. Particularly if many new companies can get to the pretraining frontier and begin to compete to push capabilities forward, and if new systems are open source, creating a low-cost or subsidized evaluations infrastructure might be essential to ensure that regulatory coverage extends beyond the large companies who can afford their own robust monitoring and compliance. Providing free evaluations in exchange for safety accreditations that could be used to attract customers or reduce insurance costs would be one useful way that regulators could try to limit avoidance of oversight by different companies. Automating evaluations, especially by integrating new forms of AI as they are produced, would be another way to limit the overall cost to regulators if the automated evaluations were of sufficient quality.

Increased regulatory capacity comes with risks of abuse but is probably fundamental to the success of any kind of regulatory regime that can respond to rapidly changing and complexifying technology. Determining how best to develop that capacity and how to deploy it in a way that is responsive to changing technology is a key part of responding to the changing frontier AI paradigm and necessary to ensure the success of other regulatory approaches that are chosen.

\section{Conclusion}

The end of the pretraining paradigm, if it comes, will present fundamental challenges for frontier AI governance. Current regulatory frameworks, built around assumptions of continued pretraining scaling, may become misaligned with the trajectory of frontier AI. Whether the pretraining frontier operates as a kind of cap on overall capabilities that deconcentrates the field and ushers in an era of more diverse and complicated innovations or incumbent players manage to find new ways to push ahead, regulations that rely on pretraining scaling will need significant adaptation. Regulators who want to preserve the virtues of efficiency and legibility that the pretraining paradigm has enabled will have to seek new ways to stay with the moving frontier. First, enhanced approaches to transparency and regulatory capacity-building will be necessary to cover a more diverse and complicated frontier field. Second, identifying whether new technical bottlenecks to AI progress emerge that can offer regulatory leverage without stifling innovation or violating rights will be a crucial part of developing effective and well-targeted regulation. This paper has sought to inform current and future regulation by laying out some initial steps forward into a new regulatory paradigm beyond the pretraining frontier that ensures that frontier AI is developed safety and in alignment with human values.

\newpage

\bibliographystyle{ACM-Reference-Format}
\bibliography{beyondbib}

\end{document}